\documentclass[aps,pre,twocolumn,showpacs,amsmath,amssymb,floatfix]{revtex4}

\usepackage[dvips]{graphics}
\usepackage{bm}

\begin{document}

\title{Quantum transport in randomly diluted quantum percolation clusters
in two dimensions}
\author{E. Cuansing}
\affiliation{Physics Department, De La Salle University, Manila 1004,
Philippines}
\author{H. Nakanishi}
\affiliation{Department of Physics, Purdue University, West Lafayette,
Indiana 47907}

\date{\today}

\begin{abstract}
We study the hopping transport of a quantum particle through finite,
randomly diluted percolation clusters in two dimensions.  We investigate
how the transmission coefficient $T$ behaves as a function of the energy
$E$ of the particle, the occupation concentration $p$ of the disordered
cluster, the size of the underlying lattice, and the type of connection
chosen between the cluster and the input and output leads.  We investigate
both the point-to-point contacts and the busbar type of connection.  For
highly diluted clusters we find the behavior of the transmission to be
independent of the type of connection.  As the amount of dilution is
decreased we find sharp variations in transmission.  These variations are
the remnants of the resonances at the ordered, zero-dilution, limit.  For
particles with energies within $0.25 \leq E \leq 1.75$ (relative to the
hopping integral) and with underlying square lattices of size $20{\times}20$,
the configurations begin transmitting near $p_{\alpha} = 0.60$ with $T$ against
$p$ curves following a common pattern as the amount of dilution is decreased.
Near $p_{\beta} = 0.90$ this pattern is broken and the transmission begins
to vary with the energy.  In the asymptotic limit of very large clusters we
find the systems to be totally reflecting except when the amount of dilution
is very low and when the particle has energy close to a resonance value at
the ordered limit or when the particle has energy at the middle of the band.

\end{abstract}

\pacs{05.60.Gg,05.50.+q,05.10.-a,73.23.-b}

\maketitle

\section{INTRODUCTION}
\label{sec:intro}

In classical systems a particle can traverse around a configuration of
barriers as long as there is an energetically available path.  However, in
systems where quantum mechanical effects cannot be ignored due to the
wavelike nature of the particle, constructive and destructive interference
can occur whenever there are paths of different lengths available to the
particle.  The transport of a particle is therefore significantly influenced
by quantum interference in systems where quantum mechanical effects cannot
be ignored.  In a previous paper \cite{cuansing04} we focused on the effects 
of quantum interference in the hopping transport of a particle through finite, 
ordered square lattices.  In this paper we extend our studies to include the
transport of a particle through finite disordered clusters in two dimensions
(2D) where the disorder is introduced by random dilution.

The question of whether disordered clusters in 2D are always insulating or 
not is controversial.  One-parameter scaling theory \cite{abrahams79} states
that there is no metallic state in non-interacting 2D disordered systems.
All states should be localized in an infinitely large and disordered 2D
system at zero temperature.  In weak localization, states are logarithmically
localized, while in strong localization, states are exponentially
localized.  In the presence of disorder, states are either weakly or
strongly localized depending on the amount of disorder.  However, results
from recent experiments on dilute low-disordered Si MOSFET and GaAs/AlGaAs
heterostructures show hints of a metallic behavior and a metal-to-insulator
transition in 2D disordered systems 
\cite{kravchenko94,kravchenko95,sarachik99}.  Several subsequent experiments
on different 2D disordered systems also produce results hinting at a
metal-to-insulator transition.  For a review of these experiments see
Abrahams {\it et al.} \cite{abrahams01}  Furthermore, experiments done on
self-assembled quantum dots on Ga[Al]As heterostructures \cite{ribeiro99}
also suggest a metal-to-insulator transition.  Since one-parameter scaling
theory deals only with non-interacting particles, it is possible that a
metallic behavior can occur as a result of the interplay between 
inter-particle interactions and the disorder.  There is also the question
of whether the type of disorder chosen is relevant.  One of the classic
and intensively investigated model of hopping transport of non-interacting
particles in disordered systems is the Anderson model of localization
\cite{anderson58}.  The particle in this model is governed by the
tight-binding Hamiltonian
\begin{equation}
  H_A = \sum_i~\epsilon_i \left|i \right> \left< i \right| 
  + \sum_{\left< ij \right>}~v_{ij} \left( \left| i \right> \left< j \right| 
  + \left| j \right> \left< i \right| \right),
\label{eq:anderson}
\end{equation}
where the $\left|i\right>$ and $\left|j\right>$ are tight-binding basis 
functions centered on sites $i$ and $j$, respectively, $v_{ij} = 1$ if $i$ 
and $j$ are nearest-neighbor sites and $v_{ij} = 0$ otherwise, and the 
on-site energies $\epsilon_i$ are randomly chosen in some range
$\left| \epsilon_i \right| \leq W$.  A particle traverses the lattice by
hopping from a site to a nearest neighbor site.  As the range $W$ is
increased a transition from conducting to localized states occurs for systems
of dimension $d \geq 3$.  However, all states are localized for systems of
dimension two or below.  The type of disorder in this model therefore can
not give rise to the observed metallic behavior in 2D disordered systems.

The disorder in Eq.~(\ref{eq:anderson}) lies in the on-site energies 
$\epsilon_i$ while the underlying lattice is ordered.  A variant of the
Anderson model is the quantum percolation model \cite{degennes59,kirkpatrick72}
wherein the on-site energies $\epsilon_i$ are held constant while the
underlying lattice is a disordered cluster from percolation theory
\cite{stauffer94}.  Since the underlying cluster is disordered not every
nearest neighbor site is available for the particle to hop onto.  In matrix
representation the disorder in the Anderson model is located along the
diagonal of the Hamiltonian while in quantum percolation the disorder is
manifest at off-diagonal locations.

There is a long-standing question whether the Anderson model and the quantum
percolation model yield the same type of localization behavior and thus,
belong to the same universality class.  For disordered clusters in three
dimensions it is widely agreed that there is a transition from extended to
localized states in quantum percolation 
\cite{soukoulis92,root86,chang95,berkovits96,chang87}.  For disordered
clusters in 2D, however, there is no clear consensus whether such a
transition exists in quantum percolation.  Several groups of researchers
including those using the dlog Pad\'{e} approximation method \cite{daboul00},
real space renormalization method \cite{odagaki84}, or studying the inverse
participation ratio \cite{srivastava84} found a transition from exponentially
localized states to non-exponentially localized states for the site
concentration that ranges within $0.73 \leq p_q \leq 0.87$.  There are,
however, studies that do not find evidence of such a transition.  Studies by
numerically calculating the conductance and making use of the one-parameter
scaling hypothesis \cite{haldas02}, by using the vibration-diffusion
analogy \cite{bunde98}, by finite-size scaling analysis and by transfer
matrix methods \cite{soukoulis91}, and vector recursion technique
\cite{mookerjee95} found no evidence of a transition.  A study by Inui 
{\it et al.} \cite{inui94} found all states to be localized except for those 
with particle energies at the middle of the band and when the underlying 
lattice is bipartite, such as a square lattice.  It is therefore not clear 
whether the Anderson model and the quantum percolation model produce the same 
type of localization behavior and if not, whether quantum percolation may be 
relevant to the experimentally observed conducting behavior in disordered 
systems in 2D.  

There are three adjustable parameters in quantum percolation: the energy $E$
of the incident particle, the concentration $p$ for the random dilution of
the lattice, and the size of the system.  In this paper we present our
studies and results by varying the values of all three parameters.

\section{COUPLING THE LEADS TO THE CLUSTER AND DETERMINING THE CONDUCTANCE}
\label{sec:coupling}

In a previous paper \cite{cuansing04} we studied the effects of quantum
interference in the transmission of a particle through ordered, i.e., 
undiluted, square lattices.  To study the transport properties we
connected semi-infinite chains to the square lattices.  The incident
particle was then set to propagate from one chain, pass through the square
lattice, and then transmit through to the other chain.  We used the 
quantum percolation model and set the constant on-site energy $\epsilon_i = 0$
in Eq.~(\ref{eq:anderson}).  Transmission and reflection coefficients were
then determined from the eigenstates of the whole system including the 
chains.  From the transmission and reflection coefficients the conductance
may be calculated by making use of the Landauer-B\"{u}ttiker formalism
\cite{buttiker85}.  For transport through ordered square lattices we found
transmission and reflection resonances whenever the energy of the incident
particle was close to a doubly-degenerate eigenvalue of the square lattice.
We also found the type of connection chosen between the chains and the
square lattice to strongly influence the transport characteristics.  Here
we follow the same method described above in determining the transport
properties of disordered clusters.

\begin{figure}[h!]
{\resizebox{3.2in}{1.35in}{\includegraphics{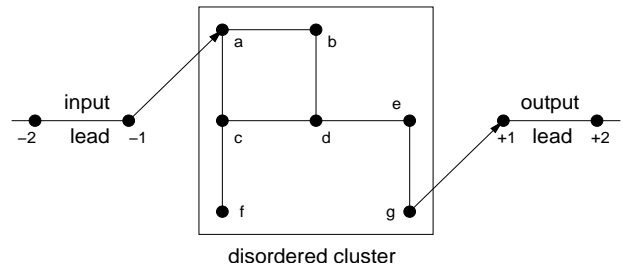}}}
\caption{An illustration of a disordered cluster on a $3{\times}3$ lattice
attached to the input and output leads through point-to-point contacts.
Beside each site is its unique label.}
\label{fig:pt2pt}
\end{figure}

\begin{figure}[h!]
{\resizebox{3.2in}{1.35in}{\includegraphics{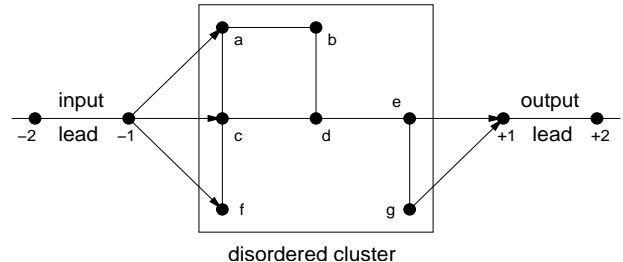}}}
\caption{An illustration of a disordered cluster on a $3{\times}3$ lattice
attached to the input and output leads through a busbar connection.
Beside each site is its unique label.  Notice that in contrast to the 
point-to-point contacts there are multiple connections between the cluster 
and the leads in a busbar.}
\label{fig:busbar}
\end{figure}

In this paper the disordered clusters are constructed using site percolation
on a square lattice \cite{stauffer94}.  There are several ways of attaching
the semi-infinite chains to the disordered clusters.  Shown in 
Figs.~\ref{fig:pt2pt} and \ref{fig:busbar} are two possible ways of 
attaching the chains.  We label the chain where the particle is incident
from as the input lead and the chain where the particle transmit through as
the output lead.  A point-to-point contacts type of connection is shown in
Fig.~\ref{fig:pt2pt}.  The input lead is attached to the top-leftmost site
while the output lead is attached to the bottom-rightmost site in the
disordered cluster.  The particle is incident through the input lead, passes
through the cluster, and then transmits through the output lead.  If the
cluster is not spanning it is not possible to complete the connection from the
input to the output lead.  The incident particle is then hindered from 
propagating through the cluster.  The minimum necessary requirement therefore
for the particle to reach the output lead is for the cluster to be spanning.
In site percolation on a square lattice the critical concentration of
occupied sites is $p_c = 0.5927$.  In quantum percolation, if there is indeed
a transition from localized to extended states as the concentration of sites
is increased then this transition should occur above the classical percolation
threshold $p_c$.

A busbar type of connection between the cluster and the leads is shown in
Fig.~\ref{fig:busbar}.  The input lead is attached to all the sites at the 
left side of the cluster while the output lead is attached to all the sites
at the right side of the cluster.  In contrast to the point-to-point contacts,
there are multiple connections between the leads and the cluster in a busbar.
A physical realization of a busbar can be a set-up wherein the contacts
connecting the disordered cluster to the current source are large enough to
encompass and attach to the whole side of the cluster.

There are other ways of connecting the disordered cluster to the input and
output leads.  We have, however, chosen the point-to-point contacts and the
busbar because at the ordered limit, when the cluster becomes a fully
occupied lattice, these two types of connections are complimentary in the
sense that point-to-point contacts maintain the bipartite symmetry of the
square lattice while the busbar, because of its multiple connections, breaks
that symmetry \cite{cuansing04}.

To determine the transport properties of the disordered clusters we follow
the method described by Daboul {\it et al.} \cite{daboul00}  Setting the 
constant on-site energy $\epsilon_i = 0$, the quantum percolation Hamiltonian 
becomes
\begin{equation}
  H_{qp} = \sum_{\left< ij \right>}~v_{ij} \left( \left| i 
  \right> \left< j \right| + \left| j \right> \left< i \right| \right).
\label{eq:qperc}
\end{equation}
We then attach the semi-infinite chains to the cluster.  The transmission
and reflection coefficients can then be determined from the resulting
eigenvalue equation $H_{qp} \psi = E \psi$.  However, this is an 
infinitely-sized problem and to reduce it to a finite one Daboul {\it et al.} 
made the following ansatz:
\begin{equation}
  \begin{array}{ccl}
    \psi_{-(n+1)} & = & e^{-inq} + r~e^{inq},\\
    \psi_{+(n+1)} & = & t~e^{inq},\\
  \end{array}
\label{eq:ansatz}
\end{equation}
where $n = 0, 1, 2, \ldots$.  Notice that shown beside each site in
Figs.~\ref{fig:pt2pt} and \ref{fig:busbar} is its unique label.  Sites
belonging to the cluster are labeled alphabetically.  Sites belonging to the
input chain are labeled by negative integers while those belonging to the
output chain are labeled by positive integers.  In the ansatz above
$\psi_{-(n+1)}$ is the part of the wavefunction along the input chain and
$\psi_{+(n+1)}$ is the part of the wavefunction along the output chain.  The
$t$ and $r$ are the transmission and reflection amplitudes, respectively.
What the ansatz implies, therefore, is that the incoming plane wave is
partially reflected back through the input chain and partially transmitted
out to the output chain.

Using the ansatz in Eq.~(\ref{eq:ansatz}) we can reduce the infinitely-sized
eigenvalue problem into a finite one.  For example, for the busbar 
configuration shown in Fig.~\ref{fig:busbar} the eigenvalue problem in
matrix representation reduces to
\begin{widetext}
\begin{equation}
  \left( \begin{array}{ccccccccc}
           -E + e^{i q} &  1 &  0 &  1 &  0 &  0 &  1 &  0 &  0 \\
                1       & -E &  1 &  1 &  0 &  0 &  0 &  0 &  0 \\
                0       &  1 & -E &  0 &  1 &  0 &  0 &  0 &  0 \\
                1       &  1 &  0 & -E &  1 &  0 &  1 &  0 &  0 \\
                0       &  0 &  1 &  1 & -E &  1 &  0 &  0 &  0 \\
                0       &  0 &  0 &  0 &  1 & -E &  0 &  1 &  1 \\
                1       &  0 &  0 &  1 &  0 &  0 & -E &  0 &  0 \\
                0       &  0 &  0 &  0 &  0 &  1 &  0 & -E &  1 \\
                0       &  0 &  0 &  0 &  0 &  1 &  0 &  1 & -E + e^{i q} \\
         \end{array} \right)
  \left( \begin{array}{c}
           1 + r \\ \psi_a \\ \psi_b \\ \psi_c \\ \psi_d \\ \psi_e \\
            \psi_f \\ \psi_g \\ t \\
         \end{array} \right) =
  \left( \begin{array}{c}
            e^{i q} - e^{-i q} \\ 0 \\ 0 \\ 0 \\ 0 \\ 0 \\ 0 \\ 0 \\ 0 \\
         \end{array} \right).
\label{eq:reduced}
\end{equation}
\end{widetext}
The connection between the cluster and the leads is manifest in the perimeter
of the $9{\times}9$ square matrix above.  The inner $7{\times}7$ inner
sub-matrix involves the connections among sites in the cluster and is
therefore independent of the type of connection between the cluster and the
leads.  In order for the ansatz to lead to valid solutions, the following
relation between $q$ and $E$ must be satisfied along the chains:
\begin{equation}
  e^{-i q} + e^{i q} = E.
\label{eq:constraint}
\end{equation}
The reflection coefficient $R = \left|r\right|^2$ and transmission coefficient
$T = \left|t\right|^2$ can then be calculated from Eqs.~(\ref{eq:reduced}) and
(\ref{eq:constraint}) and by choosing a value for the incident particle's
energy $E$.  Note that because of the constraint condition, i.e.,
Eq.~(\ref{eq:constraint}), along the chains the particle's energy is now
restricted to be within $-2 \leq E \leq 2$.  Also, although the problem now
involves a finite matrix, effectively it is still an infinite system.
Therefore, the energy $E$ is continuous and any value within the range
$\left[-2,2\right]$ leads to a valid solution.

Eq.~(\ref{eq:reduced}) is in the form of a linear equation $A x = b$ with
$x$ as the unknown.  We numerically determine $x$ exactly by solving the
inverse, $A^{-1}$, using singular value decomposition \cite{press92} and
then multiplying it to $b$.  For a given lattice size, $L{\times}L$, as we
increase the occupation concentration $p$ the size of the disordered cluster,
on average, increases as well.  Consequently, the size of the linear problem
to be solved increases.  For a disordered cluster of size $N$ the size of the
matrix $A$ to be solved is $\left(N+2\right)^2$.  In this paper the largest
clusters we investigate are those embedded in lattices of size $30{\times}30$.
We repeat the calculations of the transmission amplitude $t$ and reflection
amplitude $r$ for $1000$ realizations of the disorder and then take the
average of $\left|t\right|^2$ and $\left|r\right|^2$ to get the average
transmission coefficient $T$ and reflection coefficient $R$, respectively.

\section{NUMERICAL RESULTS}
\label{sec:numres}

\begin{figure}[h!]
{\resizebox{3.1in}{2.4in}{\includegraphics{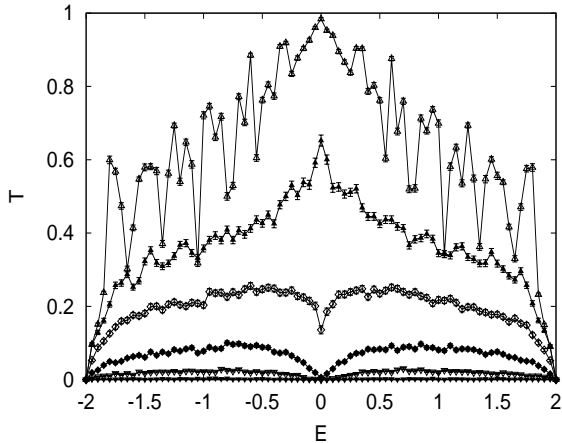}}}
\caption{Plot of the transmission coefficient $T$ as a function of the 
incident particle's energy $E$ for disordered clusters on $20{\times}20$
lattices with point-to-point contacts between the leads and the cluster.
The disorder concentrations are $p = 0.60$ ($\blacktriangledown$), $p = 0.70$ 
($\triangledown$), $p = 0.80$ ($\blacklozenge$), $p = 0.90$ ($\lozenge$), 
$p = 0.95$ ($\blacktriangle$), and $p = 0.99$ ($\vartriangle$).}
\label{fig:pt2pttvse}
\end{figure}

We investigate the characteristics of the transmission coefficient by varying
the energy $E$ of the incident particle, the site occupation concentration
$p$ of the disordered clusters, and the size $L{\times}L$ of the lattice.  
Shown in Fig.~\ref{fig:pt2pttvse} is a plot of the transmission coefficient
$T$ as a function of the energy $E$ of the incident particle.  The clusters
are connected to the input and output leads through point-to-point contacts.
In every plot shown in this paper each data point is an average over $1000$
realizations of disorder configurations.

In Fig.~\ref{fig:pt2pttvse} notice that for $p = 0.60, 0.70,$ and $0.80$ when
the clusters are highly disordered the transmission curves are shaped like
an inverted $w$ with a dip at the middle of the band.  The curves then change
shape as the occupation concentration goes from $p = 0.90$ to $p = 0.95$.  In
particular, at the middle of the band, the transmission goes from a dip to
a peak.  We also start seeing more sharp variations in the transmission
curve beginning at $p = 0.95$.  For very low amount of disorder, i.e., for
$p = 0.99$, the fluctuations are more pronounced and the peak at the middle of
the band is near the full transmission value of $T = 1$.  In our previous 
paper \cite{cuansing04} we have shown that in the case of no disorder, i.e.,
for $p = 1$, resonance occurs whenever the energy of the incident particle is
close to a doubly degenerate eigenvalue of the square lattice without the
leads.  Thus, the pronounced variations that we see in Fig.~\ref{fig:pt2pttvse}
for low amounts of disorder are actually remnants of the resonances that exist
at the ordered limit.

\begin{figure}[h!]
{\resizebox{3.1in}{2.4in}{\includegraphics{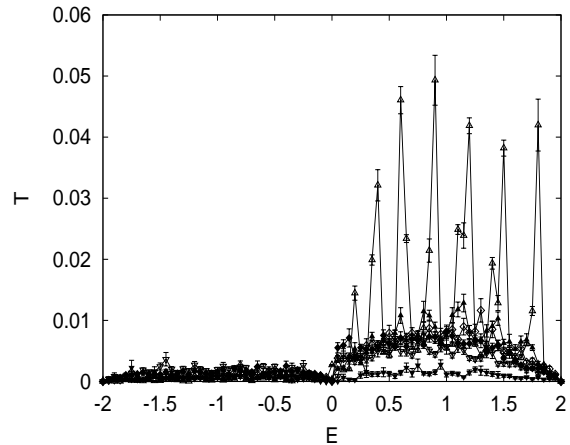}}}
\caption{Plot of the transmission coefficient $T$ against the incident
particle's energy $E$ for disordered clusters on $20{\times}20$
lattices with the busbar connection between the leads and the cluster.
The disorder concentrations are $p = 0.60$ ($\blacktriangledown$), $p = 0.70$ 
($\triangledown$), $p = 0.80$ ($\blacklozenge$), $p = 0.90$ ($\lozenge$), 
$p = 0.95$ ($\blacktriangle$), and $p = 0.99$ ($\vartriangle$).  Notice that
the transmission values are very small.}
\label{fig:busbartvse}
\end{figure}

Quantum interference occurs within the disordered cluster and also at the
connections between the leads and the cluster.  With the busbar, the multiple
connections between the cluster and the leads enhance the effect of quantum
interference.  An outcome of this is a significant decrease in the
transmission of the particle.  Shown in Fig.~\ref{fig:busbartvse} is a plot
of the transmission coefficient $T$ as a function of the energy $E$ of the
incident particle for the busbar connection.  Notice that the scale of the
transmission only goes up to $T = 0.06$.  For highly disordered clusters, just
like in the point-to-point contacts case, the shape of the transmission curve
resembles an inverted $w$ with the dip at the middle of the band.  As the
amount of disorder is decreased we see a growing asymmetry between the
positive and negative sides of $E$.  For very low amount of disorder, i.e.,
for $p = 0.99$, there are pronounced variations in the transmission.  These
variations are remnants of the resonances at the ordered limit of a square
lattice connected to leads through a busbar \cite{cuansing04}.

\begin{figure}[h!]
{\resizebox{3.1in}{2.4in}{\includegraphics{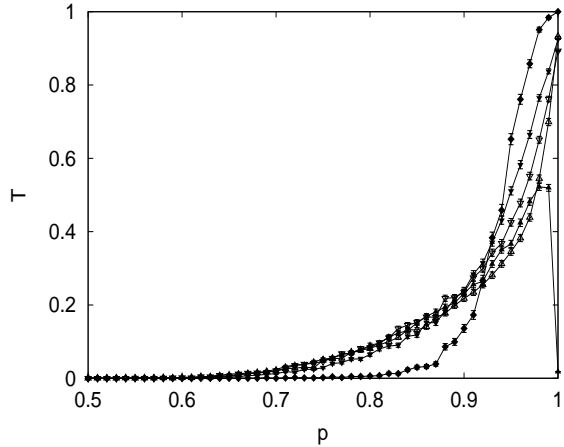}}}
\caption{Plot of the transmission coefficient $T$ as a function of the cluster
occupation concentration $p$ for disordered clusters on $20{\times}20$ 
lattices with point-to-point contacts between the leads and the cluster.  The 
incident particle has energies $E = 0.00$ ($\blacklozenge$), $E = 0.25$
($\triangle$), $E = 0.50$ ($\blacktriangle$), $E = 0.75$ ($\triangledown$),
and $E = 1.00$ ($\blacktriangledown$).}
\label{fig:pt2pttvsp}
\end{figure}

We now investigate how the transmission behaves as the disorder concentration
is varied while the energy of the particle is held fixed.  Shown in 
Fig.~\ref{fig:pt2pttvsp} is a plot of the transmission coefficient $T$ as a
function of the site occupation concentration $p$ for the point-to-point
contacts type of connection.  The results are from the $20{\times}20$
lattices.

The plot in Fig.~\ref{fig:pt2pttvsp} is for five selected values of the
particle's energy $E$.  These values are all positives since as we have seen
in Fig.~\ref{fig:pt2pttvse} there is symmetry between the positive and
negative sides of $E$ when the connection is point-to-point contacts.  In
Fig.~\ref{fig:pt2pttvsp} notice that the $E = 0$ case is special.  For high
disorder concentrations $p = 0.50$ until around $p = 0.90$ the curves for the
other $E$ values follow the same pattern.  The system begins to be transmitting
slightly above $p = 0.6$.  Near $p = 0.90$ the transmission curves begin to
spread out and at the region near $p = 1$ there are sharp variations as the
system seems to change abruptly from one dominated by disorder to one that
reflects the ordered limit wherein resonances occur.  For the system with
$E = 0$ it does not begin to transmit until around $p = 0.80$.  The
transmission curve then goes up with a slope that is steeper than the curves
for the other $E$ values.  In Fig.~\ref{fig:pt2pttvse} we have seen that at
the middle of the band the transmission shifts from being a dip to a peak as
$p$ is varied.  This shift in the behavior of the transmission is also
manifest in the rapid rise of the transmission curve after $p = 0.90$ for the
$E = 0$ case in Fig.~\ref{fig:pt2pttvsp}.

\begin{figure}[h!]
{\resizebox{3.1in}{2.4in}{\includegraphics{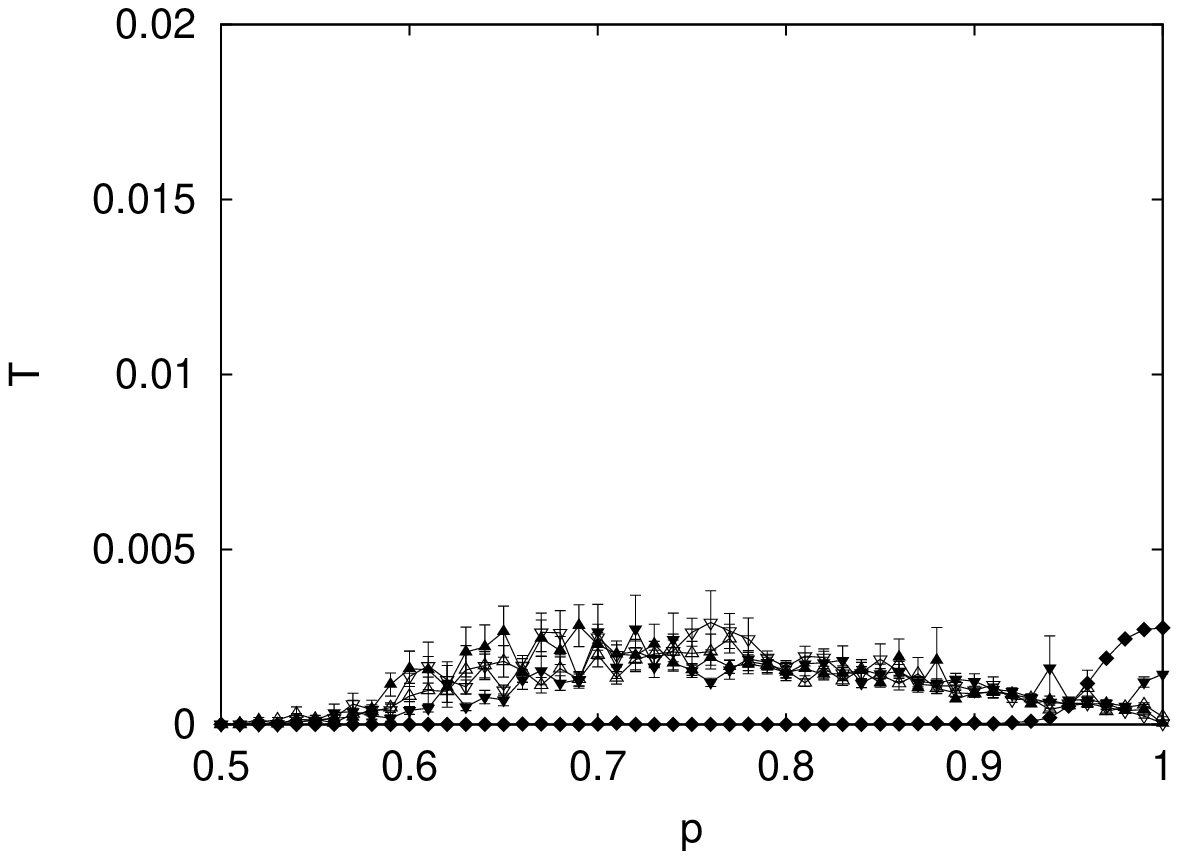}}}
\caption{Plot of the transmission coefficient $T$ as a function of the cluster
occupation concentration $p$ for disordered clusters on $20{\times}20$ 
lattices with busbar connections between the leads and the cluster.  The 
incident particle has energies $E = 0.00$ ($\blacklozenge$), $E = -0.25$
($\triangle$), $E = -0.50$ ($\blacktriangle$), $E = -0.75$ ($\triangledown$),
and $E = -1.00$ ($\blacktriangledown$).}
\label{fig:busbartvsp-}
\end{figure}

\begin{figure}[h!]
{\resizebox{3.1in}{2.4in}{\includegraphics{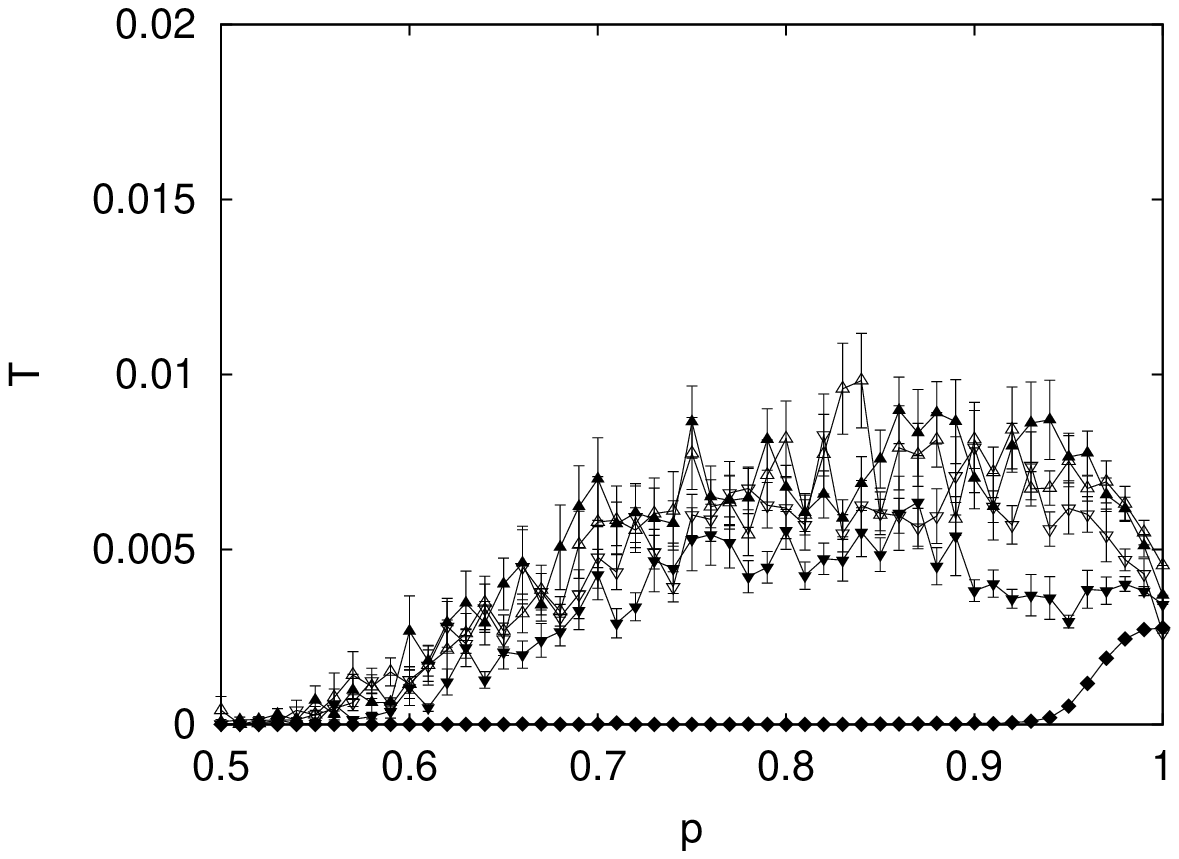}}}
\caption{Plot of the transmission coefficient $T$ as a function of the cluster
occupation concentration $p$ for disordered clusters on $20{\times}20$ 
lattices with busbar connections between the leads and the cluster.  The 
incident particle has energies $E = 0.00$ ($\blacklozenge$), $E = 0.25$
($\triangle$), $E = 0.50$ ($\blacktriangle$), $E = 0.75$ ($\triangledown$),
and $E = 1.00$ ($\blacktriangledown$).}
\label{fig:busbartvsp+}
\end{figure}

For the busbar connection, we have seen in Fig.~\ref{fig:busbartvse} that
there is no symmetry between the positive and negative sides of the energy
$E$.  We therefore show in Figs.~\ref{fig:busbartvsp-} and 
\ref{fig:busbartvsp+} plots of the transmission coefficient $T$ as a function
of the occupation concentration $p$ as the energy $E$ of the incident
particle is fixed at either negative or positive values, respectively.  The
system with the busbars is highly reflecting and we notice therefore in
Figs.~\ref{fig:busbartvsp-} and \ref{fig:busbartvsp+} that the transmission
scales are only up to $T = 0.02$.  Just like in the case for the point-to-point
contacts, the $E = 0$ case appears to be special.

In Fig.~\ref{fig:busbartvsp-} the selected energies are negative while in
Fig.~\ref{fig:busbartvsp+} the energies are positive.  The data points for
$E = 0$ are shown in both figures.  Notice again that these points do not
follow the same pattern as the data points for the other energies.  In
particular, the system is non-transmitting for high disorder concentrations
until near $p = 0.90$.  For the other $E$ values the system begins to transmit
near $p = 0.60$.  After about $p = 0.90$ the transmission curves spread out
just like in the case for point-to-point contacts.

For both the busbar and point-to-point contacts we thus see signatures of a
shift in transmission characteristics near $p_{\alpha} = 0.60$ and then again
near $p_{\beta} = 0.90$.  The system begins to transmit near $p_{\alpha}$
regardless of the type of connection chosen between the cluster and the leads,
except at the middle of the band.  Note that this value of $p$ is close to 
the classical percolation threshold $p_c$ of site percolation on a square 
lattice.  In order for the particle to transmit through a disordered cluster 
that cluster must at least be spanning.  In an infinite system, a spanning 
cluster only appears above the classical $p_c$.  However, due to quantum 
interference and the finite sizes of the systems we study, the configuration 
will not necessarily begin transmitting exactly at $p_c$.  The shift near 
$p_{\beta}$ signals the onset of the dependence of the transmission 
coefficient $T$ on the particle's energy $E$ as the occupation concentration 
$p$ is held fixed.  Between $p_{\alpha}$ and $p_{\beta}$ the transmission 
behaves in the same fashion, except, again, at the middle of the band.  This 
can also be seen in Fig.~\ref{fig:pt2pttvse} where for a high disorder 
concentration, say for $p = 0.80$, the transmission along the interval 
$0.25 < E < 1.75$ is relatively flat.

\begin{figure}[h!]
{\resizebox{3.1in}{2.4in}{\includegraphics{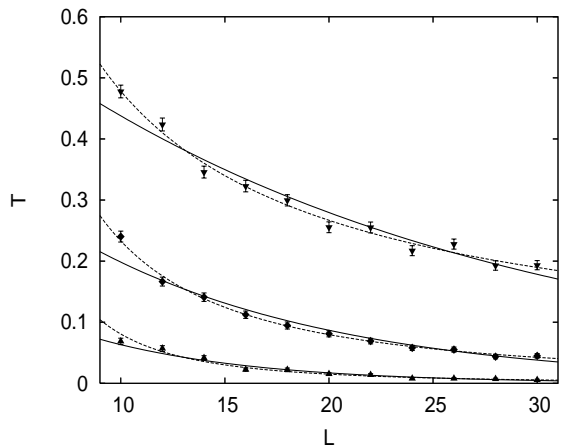}}}
\caption{Plot of the transmission coefficient $T$ as a function of the lattice
length $L$.  The energy of the incident particle is $E = 1.00$.  The
cluster and the leads are connected through point-to-point contacts.  The
occupation concentrations are $p = 0.68$ ($\blacktriangle$), $p = 0.80$ 
($\blacklozenge$), and $p = 0.92$ ($\blacktriangledown$).  The solid lines 
are the best-fitting exponential curves while the dashed lines are the 
best-fitting power-law curves.}
\label{fig:e1tvsl}
\end{figure}

\begin{figure}[h!]
{\resizebox{3.1in}{2.4in}{\includegraphics{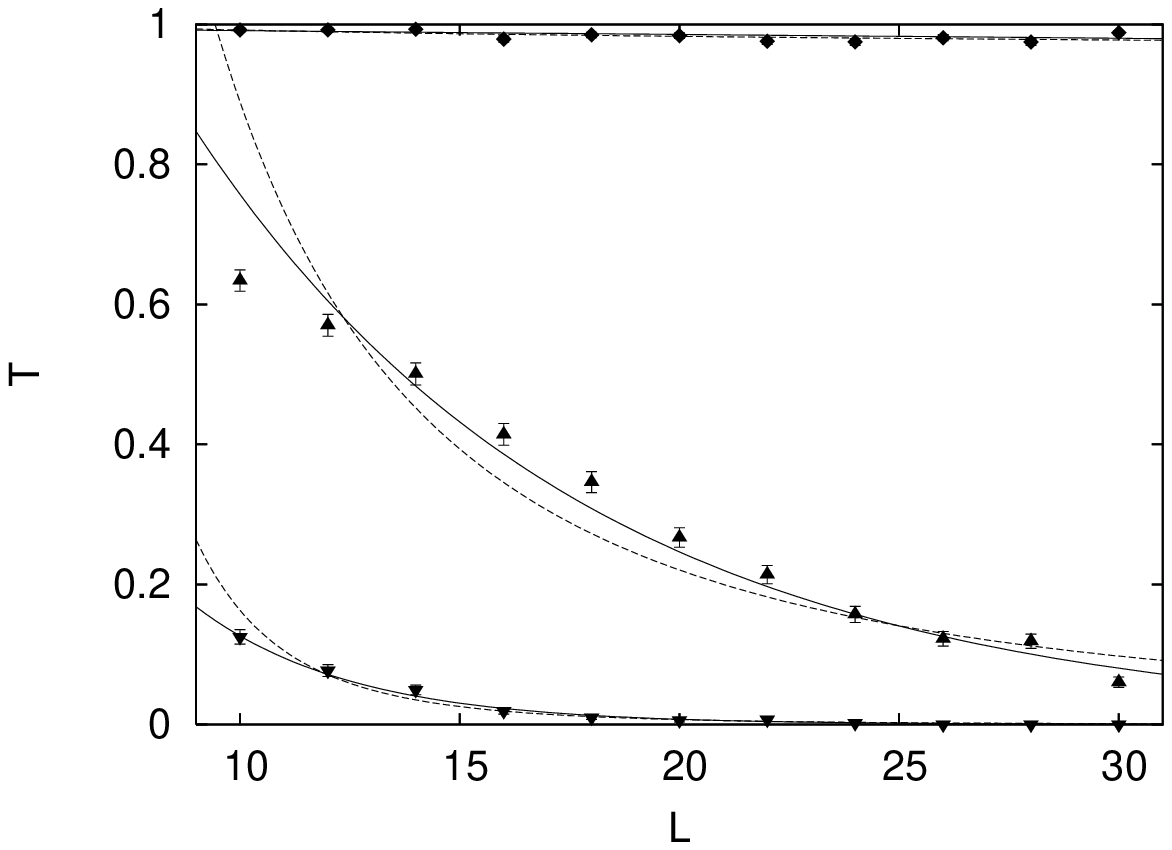}}}
\caption{The transmission coefficient $T$ as a function of the lattice length 
$L$ with an incident particle of energy $E = 0.00$.  The occupation
concentrations are $p = 0.80$ ($\blacktriangledown$), $p = 0.92$
($\blacktriangle$) and $p = 0.99$ ($\blacklozenge$).  The solid lines are the 
best-fitting exponential curves while the dashed lines are the best-fitting 
power-law curves.  For $p = 0.80$ the three points after $L = 25$ have 
transmissions that are practically zeroes and are therefore not included when 
the best-fit lines are determined.}
\label{fig:e0tvsl}
\end{figure}

We further investigate the characteristics of the transmission as the size
of the lattices are varied.  Shown in Figs.~\ref{fig:e1tvsl} and
\ref{fig:e0tvsl} are plots of the transmission as a function of lattice
sizes as the occupation concentration $p$ and energy $E$ of the incident
particle are both held fixed.  Also shown in the plots are the best-fitting
exponential curves and best-fitting power-law curves.

In Fig.~\ref{fig:e1tvsl} we have chosen the energy of the particle as $E = 1$.
We have also chosen the connection between the cluster and the leads to be
point-to-point contacts because this connection type is generally more
transmitting than the busbar.  The cluster occupation concentrations are 
chosen as $p = 0.68$ and $p = 0.92$ since these values are near $p_{\alpha}$
and $p_{\beta}$.  The power-law fits are in the form $T = T_p L^{-\sigma_p}$
while the exponential fits are in the form $T = T_e e^{-\sigma_e L}$, for 
lattices of size $L{\times}L$.  Shown in Table~\ref{tab:e1fit} are the
fitting values corresponding to the best-fit curves in Fig.~\ref{fig:e1tvsl}.
The $R_p$ and $R_e$ are the correlation coefficients for the power-law and
exponential fits performed as linear regressions of ${\log}\left(T\right)$
against ${\log}\left(L\right)$ and ${\log}\left(T\right)$ against $L$, 
respectively.

\begin{table}[h!]
  \begin{center}
  \begin{tabular}{||c||c|c|c||c|c|c||}
    \hline
    $p$ & $T_p$ & $\sigma_p$ & $\left| R_p \right|^2$ & $T_e$ & $\sigma_e$ & 
    $\left| R_e \right|^2$ \\
    \hline
    $0.68$ & $21.67$ & $2.429$ & $0.97911$ & $0.234$ & $0.131$ & $0.97001$ \\
    $0.80$ & $8.249$ & $1.549$ & $0.99455$ & $0.453$ & $0.083$ & $0.96994$ \\
    $0.92$ & $3.323$ & $0.842$ & $0.98331$ & $0.686$ & $0.045$ & $0.95684$ \\
    \hline
  \end{tabular}
  \end{center}
\caption{Table of fitting values for the $E = 1$ case.  The $T_p$, $\sigma_p$, 
and $R_p$ belong to the power-law fits while the $T_e$, $\sigma_e$, and $R_e$ 
are for the exponential fits.}
\label{tab:e1fit}
\end{table}

From the values of the correlation coefficients, $R_p$ and $R_e$, it is not
possible to clearly distinguish between the goodness of the power-law and
exponential fits as they are very close to each other.  However, since these
correlation coefficients are reasonably close to $1$, we can be fairly
confident that $T$ extrapolates to zero as $L \rightarrow \infty$ or in the
thermodynamic limit.  Thus, in either case, our result for $E = 1$ is
consistent with one-parameter scaling theory \cite{abrahams79} which predicts
that wavefunctions in two dimensions are localized.

Shown in Fig.~\ref{fig:e0tvsl} is a plot of $T$ as a function of $L$ when the
particle's energy is $E = 0$.  The corresponding best-fit estimates are shown
in Table~\ref{tab:e0fit}.  While the power-law and exponential curves are both
reasonably good for $p = 0.80$, the power-law fit is not nearly as reasonable
for $p = 0.92$.  For very low amount of disorder at $p = 0.99$ neither the
power-law nor the exponential fit is reasonable.  From the values of 
$\sigma_{0p}$ and $\sigma_{0e}$ we get a hint that the $T$ can be independent 
of $L$ at very low disorder concentrations, at least at certain values of $E$.
For such a situation, it is possible for the transmission to be non-zero even 
for asymptotically large clusters.

\begin{table}[h!]
  \begin{center}
  \begin{tabular}{||c||c|c|c||c|c|c||}
    \hline
    $p$ & $T_{0 p}$ & $\sigma_{0 p}$ & $\left| R_{0 p} \right|^2$ & 
    $T_{0 e}$ & $\sigma_{0 e}$ & $\left| R_{0 e} \right|^2$ \\
    \hline
    $0.80$ & $6122$ & $4.575$ & $0.95824$ & $2.185$ & $0.285$ & $0.96564$ \\
    $0.92$ & $91.2$ & $2.011$ & $0.90850$ & $2.325$ & $0.112$ & $0.96878$ \\
    $0.99$ & $1.02$ & $0.013$ & $0.44440$ & $0.997$ & $0.001$ & $0.37233$ \\
    \hline
  \end{tabular}
  \end{center}
\caption{Table of fitting values for the $E = 0$ case.  The $T_{0 p}$, 
$\sigma_{0 p}$, and $R_{0 p}$ belong to the power-law fits while the 
$T_{0 e}$, $\sigma_{0 e}$, and $R_{0 e}$ are for the exponential fits.  Note 
that neither fits are good for the $p = 0.99$ case.}
\label{tab:e0fit}
\end{table}

\section{SUMMARY AND CONCLUSIONS}
\label{sec:summary}

The input and output leads can be attached to the disordered cluster in
several different ways.  In this paper we have chosen the connections to be
point-to-point contacts and the busbar.  The behavior of the transmission $T$
is independent of the type of connection chosen for highly disordered
clusters.  This is because the number of actual multiple connections in the
busbar case is not large enough to differentiate its effects from that of the
point-to-point contacts type of connection.  As the site dilution is 
diminished, i.e., as $p$ increases, the effects of the choice of connection 
begin to be apparent.  For low disorder, i.e., high $p$, point-to-point 
contacts, in general, are more transmitting than the busbar.  The resonances 
at the ordered limit, however, influence the transmission at low disorder.  We 
thus see transmission variations near $p = 1$ as we vary the energy of the
particle.

We have shown how the transmission coefficient $T$ behaves as a function of
the energy $E$ of the particle, the occupation concentration $p$ of the
disordered cluster, and the size $L{\times}L$ of the underlying lattice.
For $20{\times}20$ lattices we find the system to begin transmitting around
$p_{\alpha} = 0.60$, which is near the classical percolation threshold $p_c$.
The transmission is independent of the incident particle's energy along
the interval $0.25 \leq E \leq 1.75$ when $p$ is increased until around
$p_{\beta} = 0.90$.  After $p_{\beta}$ the transmission begins to be
dependent on the energy of the particle.  The exception to this pattern is
when the particle has energy at the middle of the band.  For such a particle
the system does not begin to be transmitting until after around $p = 0.80$
in point-to-point contacts and $p = 0.90$ in the busbar.

Extrapolating the behavior of the transmission to very large lattices, we
find $T \rightarrow 0$ in moderate to high disorder.  The system is localized
in these situations.  The behavior of the transmission at very low disorder,
however, is not as straightforward.  The resonances at the ordered limit
influences the behavior of the transmission at very low disorder.  So even
with asymptotically large clusters, as long as the disorder is low, it may
be possible to get non-zero transmission when the energy of the incident
particle is close to the resonances at the ordered limit.  When the energy
of the incident particle is at the middle of the band, the system can also be
transmitting at very low disorder even for asymptotically large clusters.
This case is consistent with the work of Inui {\it et al.} \cite{inui94} where
they cite the bipartite symmetry of the underlying square lattice as the reason
for finding wavefunctions that are not exponentially localized at the middle
of the band.

Quantum percolation in two dimensions in asymptotically large clusters 
therefore generally lead to non-conducting systems.  Our results suggest that
the only possible exceptions occur at very low disorder when the incident
particle's energy is either at the middle of the band or near the resonance
value in the ordered limit.  Quantum percolation is a single-particle
quantum model of hopping transport where the effects from mechanisms such as
tunneling, long-ranged hopping, or inter-particle interactions are not taken
into account.  Taking these other effects into account may enhance the
transmission of the particle in such a way that the configuration may be
transmitting even for asymptotically large and highly disordered clusters.

\acknowledgments

We would like to thank the Rosen Center for Advanced Computing and one of
us (EC) would also like to thank the Department of Physics at Purdue 
University where most of this work was done.  We are also grateful to 
N. Giordano, A. Hirsch and I. Szleifer for illuminating discussions.

\bibliography{references}

\begin{thebibliography}{26}
\expandafter\ifx\csname natexlab\endcsname\relax\def\natexlab#1{#1}\fi
\expandafter\ifx\csname bibnamefont\endcsname\relax
  \def\bibnamefont#1{#1}\fi
\expandafter\ifx\csname bibfnamefont\endcsname\relax
  \def\bibfnamefont#1{#1}\fi
\expandafter\ifx\csname citenamefont\endcsname\relax
  \def\citenamefont#1{#1}\fi
\expandafter\ifx\csname url\endcsname\relax
  \def\url#1{\texttt{#1}}\fi
\expandafter\ifx\csname urlprefix\endcsname\relax\def\urlprefix{URL }\fi
\providecommand{\bibinfo}[2]{#2}
\providecommand{\eprint}[2][]{\url{#2}}

\bibitem[{\citenamefont{Cuansing and Nakanishi}(2004)}]{cuansing04}
\bibinfo{author}{\bibfnamefont{E.}~\bibnamefont{Cuansing}} \bibnamefont{and}
  \bibinfo{author}{\bibfnamefont{H.}~\bibnamefont{Nakanishi}},
  \bibinfo{journal}{Phys.\ Rev.\ E} \textbf{\bibinfo{volume}{70}},
  \bibinfo{pages}{066142} (\bibinfo{year}{2004}).

\bibitem[{\citenamefont{Abrahams et~al.}(1979)\citenamefont{Abrahams, Anderson,
  Licciardello, and Ramakrishnan}}]{abrahams79}
\bibinfo{author}{\bibfnamefont{E.}~\bibnamefont{Abrahams}},
  \bibinfo{author}{\bibfnamefont{P.~W.} \bibnamefont{Anderson}},
  \bibinfo{author}{\bibfnamefont{D.~C.} \bibnamefont{Licciardello}},
  \bibnamefont{and} \bibinfo{author}{\bibfnamefont{T.~V.}
  \bibnamefont{Ramakrishnan}}, \bibinfo{journal}{Phys.\ Rev.\ Lett.}
  \textbf{\bibinfo{volume}{42}}, \bibinfo{pages}{673} (\bibinfo{year}{1979}).

\bibitem[{\citenamefont{Kravchenko et~al.}(1994)\citenamefont{Kravchenko,
  Kravchenko, Furneaux, Pudalov, and D'Iorio}}]{kravchenko94}
\bibinfo{author}{\bibfnamefont{S.~V.} \bibnamefont{Kravchenko}},
  \bibinfo{author}{\bibfnamefont{G.~V.} \bibnamefont{Kravchenko}},
  \bibinfo{author}{\bibfnamefont{J.~E.} \bibnamefont{Furneaux}},
  \bibinfo{author}{\bibfnamefont{V.~M.} \bibnamefont{Pudalov}},
  \bibnamefont{and} \bibinfo{author}{\bibfnamefont{M.}~\bibnamefont{D'Iorio}},
  \bibinfo{journal}{Phys.\ Rev.\ B} \textbf{\bibinfo{volume}{50}},
  \bibinfo{pages}{8039} (\bibinfo{year}{1994}).

\bibitem[{\citenamefont{Kravchenko et~al.}(1995)\citenamefont{Kravchenko,
  Mason, Bowker, Furneaux, Pudalov, and D'Iorio}}]{kravchenko95}
\bibinfo{author}{\bibfnamefont{S.~V.} \bibnamefont{Kravchenko}},
  \bibinfo{author}{\bibfnamefont{W.~E.} \bibnamefont{Mason}},
  \bibinfo{author}{\bibfnamefont{G.~E.} \bibnamefont{Bowker}},
  \bibinfo{author}{\bibfnamefont{J.~E.} \bibnamefont{Furneaux}},
  \bibinfo{author}{\bibfnamefont{V.~M.} \bibnamefont{Pudalov}},
  \bibnamefont{and} \bibinfo{author}{\bibfnamefont{M.}~\bibnamefont{D'Iorio}},
  \bibinfo{journal}{Phys.\ Rev.\ B} \textbf{\bibinfo{volume}{51}},
  \bibinfo{pages}{7038} (\bibinfo{year}{1995}).

\bibitem[{\citenamefont{Sarachik and Kravchenko}(1999)}]{sarachik99}
\bibinfo{author}{\bibfnamefont{M.~P.} \bibnamefont{Sarachik}} \bibnamefont{and}
  \bibinfo{author}{\bibfnamefont{S.~V.} \bibnamefont{Kravchenko}},
  \bibinfo{journal}{Proc.\ Natl.\ Acad.\ Sci.\ U.S.A.}
  \textbf{\bibinfo{volume}{96}}, \bibinfo{pages}{5900} (\bibinfo{year}{1999}).

\bibitem[{\citenamefont{Abrahams et~al.}(2001)\citenamefont{Abrahams,
  Kravchenko, and Sarachik}}]{abrahams01}
\bibinfo{author}{\bibfnamefont{E.}~\bibnamefont{Abrahams}},
  \bibinfo{author}{\bibfnamefont{S.~V.} \bibnamefont{Kravchenko}},
  \bibnamefont{and} \bibinfo{author}{\bibfnamefont{M.~P.}
  \bibnamefont{Sarachik}}, \bibinfo{journal}{Rev.\ Mod.\ Phys.}
  \textbf{\bibinfo{volume}{73}}, \bibinfo{pages}{251} (\bibinfo{year}{2001}).

\bibitem[{\citenamefont{Ribeiro et~al.}(1999)\citenamefont{Ribeiro, J\"{a}ggi,
  Heinzel, and Ensslin}}]{ribeiro99}
\bibinfo{author}{\bibfnamefont{E.}~\bibnamefont{Ribeiro}},
  \bibinfo{author}{\bibfnamefont{R.~D.} \bibnamefont{J\"{a}ggi}},
  \bibinfo{author}{\bibfnamefont{T.}~\bibnamefont{Heinzel}}, \bibnamefont{and}
  \bibinfo{author}{\bibfnamefont{K.}~\bibnamefont{Ensslin}},
  \bibinfo{journal}{Phys.\ Rev.\ Lett.} \textbf{\bibinfo{volume}{82}},
  \bibinfo{pages}{996} (\bibinfo{year}{1999}).

\bibitem[{\citenamefont{Anderson}(1958)}]{anderson58}
\bibinfo{author}{\bibfnamefont{P.~W.} \bibnamefont{Anderson}},
  \bibinfo{journal}{Phys.\ Rev.} \textbf{\bibinfo{volume}{109}},
  \bibinfo{pages}{1492} (\bibinfo{year}{1958}).

\bibitem[{\citenamefont{de~Gennes et~al.}(1959)\citenamefont{de~Gennes, Lafore,
  and Millot}}]{degennes59}
\bibinfo{author}{\bibfnamefont{P.~G.} \bibnamefont{de~Gennes}},
  \bibinfo{author}{\bibfnamefont{P.}~\bibnamefont{Lafore}}, \bibnamefont{and}
  \bibinfo{author}{\bibfnamefont{J.}~\bibnamefont{Millot}},
  \bibinfo{journal}{J.\ Phys.\ Chem.\ Solids} \textbf{\bibinfo{volume}{11}},
  \bibinfo{pages}{105} (\bibinfo{year}{1959}).

\bibitem[{\citenamefont{Kirkpatrick and Eggarter}(1972)}]{kirkpatrick72}
\bibinfo{author}{\bibfnamefont{S.}~\bibnamefont{Kirkpatrick}} \bibnamefont{and}
  \bibinfo{author}{\bibfnamefont{T.~P.} \bibnamefont{Eggarter}},
  \bibinfo{journal}{Phys.\ Rev.\ B} \textbf{\bibinfo{volume}{6}},
  \bibinfo{pages}{3598} (\bibinfo{year}{1972}).

\bibitem[{\citenamefont{Stauffer and Aharony}(1994)}]{stauffer94}
\bibinfo{author}{\bibfnamefont{D.}~\bibnamefont{Stauffer}} \bibnamefont{and}
  \bibinfo{author}{\bibfnamefont{A.}~\bibnamefont{Aharony}},
  \emph{\bibinfo{title}{Introduction to Percolation Theory}}
  (\bibinfo{publisher}{Taylor \& Francis}, \bibinfo{address}{Bristol, PA},
  \bibinfo{year}{1994}), \bibinfo{edition}{2nd} ed.

\bibitem[{\citenamefont{Soukoulis et~al.}(1992)\citenamefont{Soukoulis, Li, and
  Grest}}]{soukoulis92}
\bibinfo{author}{\bibfnamefont{C.~M.} \bibnamefont{Soukoulis}},
  \bibinfo{author}{\bibfnamefont{Q.}~\bibnamefont{Li}}, \bibnamefont{and}
  \bibinfo{author}{\bibfnamefont{G.~S.} \bibnamefont{Grest}},
  \bibinfo{journal}{Phys.\ Rev.\ B} \textbf{\bibinfo{volume}{45}},
  \bibinfo{pages}{7724} (\bibinfo{year}{1992}).

\bibitem[{\citenamefont{Root and Skinner}(1986)}]{root86}
\bibinfo{author}{\bibfnamefont{L.}~\bibnamefont{Root}} \bibnamefont{and}
  \bibinfo{author}{\bibfnamefont{J.~L.} \bibnamefont{Skinner}},
  \bibinfo{journal}{Phys.\ Rev.\ B} \textbf{\bibinfo{volume}{33}},
  \bibinfo{pages}{7738} (\bibinfo{year}{1986}).

\bibitem[{\citenamefont{Chang et~al.}(1995)\citenamefont{Chang, Lev, Harris,
  Adler, and Aharony}}]{chang95}
\bibinfo{author}{\bibfnamefont{I.}~\bibnamefont{Chang}},
  \bibinfo{author}{\bibfnamefont{Z.}~\bibnamefont{Lev}},
  \bibinfo{author}{\bibfnamefont{A.~B.} \bibnamefont{Harris}},
  \bibinfo{author}{\bibfnamefont{J.}~\bibnamefont{Adler}}, \bibnamefont{and}
  \bibinfo{author}{\bibfnamefont{A.}~\bibnamefont{Aharony}},
  \bibinfo{journal}{Phys.\ Rev.\ Lett.} \textbf{\bibinfo{volume}{74}},
  \bibinfo{pages}{2094} (\bibinfo{year}{1995}).

\bibitem[{\citenamefont{Berkovits and Avishai}(1996)}]{berkovits96}
\bibinfo{author}{\bibfnamefont{R.}~\bibnamefont{Berkovits}} \bibnamefont{and}
  \bibinfo{author}{\bibfnamefont{Y.}~\bibnamefont{Avishai}},
  \bibinfo{journal}{Phys.\ Rev.\ B} \textbf{\bibinfo{volume}{53}},
  \bibinfo{pages}{R16125} (\bibinfo{year}{1996}).

\bibitem[{\citenamefont{Chang and Odagaki}(1987)}]{chang87}
\bibinfo{author}{\bibfnamefont{K.~C.} \bibnamefont{Chang}} \bibnamefont{and}
  \bibinfo{author}{\bibfnamefont{T.}~\bibnamefont{Odagaki}},
  \bibinfo{journal}{J.\ Phys.\ A:Math.\ Gen.} \textbf{\bibinfo{volume}{20}},
  \bibinfo{pages}{L1027} (\bibinfo{year}{1987}).

\bibitem[{\citenamefont{Daboul et~al.}(2000)\citenamefont{Daboul, Chang, and
  Aharony}}]{daboul00}
\bibinfo{author}{\bibfnamefont{D.}~\bibnamefont{Daboul}},
  \bibinfo{author}{\bibfnamefont{I.}~\bibnamefont{Chang}}, \bibnamefont{and}
  \bibinfo{author}{\bibfnamefont{A.}~\bibnamefont{Aharony}},
  \bibinfo{journal}{Eur.\ Phys.\ J.\ B} \textbf{\bibinfo{volume}{16}},
  \bibinfo{pages}{303} (\bibinfo{year}{2000}).

\bibitem[{\citenamefont{Odagaki and Chang}(1984)}]{odagaki84}
\bibinfo{author}{\bibfnamefont{T.}~\bibnamefont{Odagaki}} \bibnamefont{and}
  \bibinfo{author}{\bibfnamefont{K.~C.} \bibnamefont{Chang}},
  \bibinfo{journal}{Phys.\ Rev.\ B} \textbf{\bibinfo{volume}{30}},
  \bibinfo{pages}{1612} (\bibinfo{year}{1984}).

\bibitem[{\citenamefont{Srivastava and Chaturvedi}(1984)}]{srivastava84}
\bibinfo{author}{\bibfnamefont{V.}~\bibnamefont{Srivastava}} \bibnamefont{and}
  \bibinfo{author}{\bibfnamefont{M.}~\bibnamefont{Chaturvedi}},
  \bibinfo{journal}{Phys.\ Rev.\ B} \textbf{\bibinfo{volume}{30}},
  \bibinfo{pages}{2238} (\bibinfo{year}{1984}).

\bibitem[{\citenamefont{Ha{\l}da{\'s} et~al.}(2002)\citenamefont{Ha{\l}da{\'s},
  Kolek, and Stadler}}]{haldas02}
\bibinfo{author}{\bibfnamefont{G.}~\bibnamefont{Ha{\l}da{\'s}}},
  \bibinfo{author}{\bibfnamefont{A.}~\bibnamefont{Kolek}}, \bibnamefont{and}
  \bibinfo{author}{\bibfnamefont{A.~W.} \bibnamefont{Stadler}},
  \bibinfo{journal}{Phys.\ Status Solidi B} \textbf{\bibinfo{volume}{230}},
  \bibinfo{pages}{249} (\bibinfo{year}{2002}).

\bibitem[{\citenamefont{Bunde et~al.}(1998)\citenamefont{Bunde, Kantelhardt,
  and Schweitzer}}]{bunde98}
\bibinfo{author}{\bibfnamefont{A.}~\bibnamefont{Bunde}},
  \bibinfo{author}{\bibfnamefont{J.~W.} \bibnamefont{Kantelhardt}},
  \bibnamefont{and}
  \bibinfo{author}{\bibfnamefont{L.}~\bibnamefont{Schweitzer}},
  \bibinfo{journal}{Ann.\ Phys.\ (Leipzig)} \textbf{\bibinfo{volume}{7}},
  \bibinfo{pages}{372} (\bibinfo{year}{1998}).

\bibitem[{\citenamefont{Soukoulis and Grest}(1991)}]{soukoulis91}
\bibinfo{author}{\bibfnamefont{C.~M.} \bibnamefont{Soukoulis}}
  \bibnamefont{and} \bibinfo{author}{\bibfnamefont{G.~S.} \bibnamefont{Grest}},
  \bibinfo{journal}{Phys.\ Rev.\ B} \textbf{\bibinfo{volume}{44}},
  \bibinfo{pages}{4685} (\bibinfo{year}{1991}).

\bibitem[{\citenamefont{Mookerjee et~al.}(1995)\citenamefont{Mookerjee,
  Dasgupta, and Saha}}]{mookerjee95}
\bibinfo{author}{\bibfnamefont{A.}~\bibnamefont{Mookerjee}},
  \bibinfo{author}{\bibfnamefont{I.}~\bibnamefont{Dasgupta}}, \bibnamefont{and}
  \bibinfo{author}{\bibfnamefont{T.}~\bibnamefont{Saha}},
  \bibinfo{journal}{Int.\ J.\ Mod.\ Phys.\ B} \textbf{\bibinfo{volume}{9}},
  \bibinfo{pages}{2989} (\bibinfo{year}{1995}).

\bibitem[{\citenamefont{Inui et~al.}(1994)\citenamefont{Inui, Trugman, and
  Abrahams}}]{inui94}
\bibinfo{author}{\bibfnamefont{M.}~\bibnamefont{Inui}},
  \bibinfo{author}{\bibfnamefont{S.~A.} \bibnamefont{Trugman}},
  \bibnamefont{and} \bibinfo{author}{\bibfnamefont{E.}~\bibnamefont{Abrahams}},
  \bibinfo{journal}{Phys.\ Rev.\ B} \textbf{\bibinfo{volume}{49}},
  \bibinfo{pages}{3190} (\bibinfo{year}{1994}).

\bibitem[{\citenamefont{B\"{u}ttiker et~al.}(1985)\citenamefont{B\"{u}ttiker,
  Imry, Landauer, and Pinhas}}]{buttiker85}
\bibinfo{author}{\bibfnamefont{M.}~\bibnamefont{B\"{u}ttiker}},
  \bibinfo{author}{\bibfnamefont{Y.}~\bibnamefont{Imry}},
  \bibinfo{author}{\bibfnamefont{R.}~\bibnamefont{Landauer}}, \bibnamefont{and}
  \bibinfo{author}{\bibfnamefont{S.}~\bibnamefont{Pinhas}},
  \bibinfo{journal}{Phys.\ Rev.\ B} \textbf{\bibinfo{volume}{31}},
  \bibinfo{pages}{6207} (\bibinfo{year}{1985}).

\bibitem[{\citenamefont{Press et~al.}(1992)\citenamefont{Press, Teukolsky,
  Vetterling, and Flannery}}]{press92}
\bibinfo{author}{\bibfnamefont{W.}~\bibnamefont{Press}},
  \bibinfo{author}{\bibfnamefont{S.}~\bibnamefont{Teukolsky}},
  \bibinfo{author}{\bibfnamefont{W.}~\bibnamefont{Vetterling}},
  \bibnamefont{and} \bibinfo{author}{\bibfnamefont{B.}~\bibnamefont{Flannery}},
  \emph{\bibinfo{title}{Numerical Recipes in Fortran}}
  (\bibinfo{publisher}{Cambridge University Press},
  \bibinfo{address}{Cambridge, U.K.}, \bibinfo{year}{1992}),
  \bibinfo{edition}{2nd} ed.

\end{thebibliography}

\end{document}